\documentclass[preprint,12pt,a4paper]{elsarticle}

\usepackage{amssymb}

\journal{J. Solid State Chem.}

\begin{document}

\begin{frontmatter}

\title{Competition between ferromagnetic and antiferromagnetic states in Al$_{8.5-x}$Fe$_{23}$Ge$_{12.5+x}$ (0$\leq$$x$$\leq$3)}

\author{Jiro Kitagawa$^1$}

\address{$^1$ Department of Electrical Engineering, Faculty of Engineering, Fukuoka Institute of Technology, 3-30-1 Wajiro-higashi, Higashi-ku, Fukuoka 811-0295, Japan}
\ead{j-kitagawa@fit.ac.jp}

\author{Genta Yakabe$^1$}

\address{$^1$ Department of Electrical Engineering, Faculty of Engineering, Fukuoka Institute of Technology, 3-30-1 Wajiro-higashi, Higashi-ku, Fukuoka 811-0295, Japan}

\author{Akinori Nakayama$^1$}

\address{$^1$ Department of Electrical Engineering, Faculty of Engineering, Fukuoka Institute of Technology, 3-30-1 Wajiro-higashi, Higashi-ku, Fukuoka 811-0295, Japan}

\author{Terukazu Nishizaki$^2$}

\address{$^2$ Department of Electrical Engineering, Faculty of Science and Engineering, Kyushu Sangyo University, 2-3-1 Matsukadai, Higashi-ku, Fukuoka 813-8503, Japan}

\author{Masami Tsubota$^3$}

\address{$^3$ Physonit Inc., 6-10 Minami-Horikawa, Kaita Aki, Hiroshima 736-0044, Japan}

\begin{abstract}
Polycrystalline Al$_{7+x}$Fe$_{23}$Ge$_{14-x}$-type orthorhombic Al$_{8.5-x}$Fe$_{23}$Ge$_{12.5+x}$ (0$\leq$$x$$\leq$3) compounds were studied by X-ray diffraction measurement, metallographic examination, and magnetization and electrical resistivity measurements. The lattice parameters show anisotropic $x$ dependences, reflecting a lower symmetry of rather complex crystal structure. In the $x$=0 sample, a ferromagnetic (FM) transition occurs at the Curie temperature $T_\mathrm{C}$ of 255 K, which is systematically reduced with increasing $x$ and disappears at $x$=2.5 where an antiferromagnetic (AFM) ground state is realized. The N\'eel temperature $T_\mathrm{N}$ seems to appear at $x$$\geq$1, and grows to 143 K at $x$=3 with the increment of $x$. In the intermediate range of $x$ (1$\leq$$x$$\leq$2), a FM to AFM phase transition occurs below $T_\mathrm{C}$. In the AFM phase, the electrical resistivity $\rho$ shows a Fermi surface instability characterized by the upturn of $\rho$ below approximately $T_\mathrm{N}$. The competition between the FM and AFM states would be ascribed to the anisotropic $x$ dependences of Fe-Fe bond lengths.
\end{abstract}


\end{frontmatter}

\clearpage

\section{Introduction}
An intermetallic compound with a complex crystal structure, possessing many crystallographic sites and/or a large number of atoms in the unit cell, often shows an interesting phenomenon as follows.
Ce$_{3}$Pd$_{20}$Si(Ge)$_{6}$ and LaFe$_{13-x}$Si$_{x}$ crystallize into the cubic C$_{6}$Cr$_{23}$-type structure ({\it Fm}$\bar{3}${\it m}, No. 225, cF116) and the cubic NaZn$_{13}$-type structure ({\it Fm}$\bar{3}${\it c}, No. 226, cF112), respectively.
The former compound exhibits the coexistence of the Kondo effect and the quadrupolar ordering\cite{Takeda:JPSP1995,Kitagawa:PRB1996,Kitagawa:JAC1997,Kitagawa:PRB1998,Kitagawa:JPSJ2000,Custers:NM2012}, and the latter one does a high magnetic refrigeration efficiency\cite{Fujita:PRB2003}.
The Th$_{2}$Zn$_{17}$-type ({\it R}$\bar{3}${\it m}, No. 166-2, hP57) and Nd$_{2}$Fe$_{14}$B-type ({\it P}4$_{2}$/{\it mnm}, No. 136, tP68) structures are famous for the permanent magnets Sm$_{2}$Co$_{17}$, Sm$_{2}$Fe$_{17}$N$_{x}$ , Nd$_{2}$Fe$_{14}$B and so on\cite{Tawara:JJAP1973,Sagawa:JAP1984,Croaf:JAP1984,Coey:JMMM1990}, and the Fe atoms of both structures have four crystallographically independent sites.
Yb$_{14}$MnSb$_{11}$ with the tetragonal Ca$_{14}$AlSb$_{11}$-type ({\it I}4$_{1}$/{\it acd}, No. 142, tI208) structure attracts much attention as a good candidate showing a high thermoelectric figure of merit\cite{Brown:ChemMater2006}.

We have focused on Al$_{8.5-x}$Fe$_{23}$Ge$_{12.5+x}$, which has been recently discovered and crystallizes into a complex orthorhombic Al$_{7+x}$Fe$_{23}$Ge$_{14-x}$-type structure ({\it Cmc}2$_{1}$, No. 36, oC176)\cite{Reisinger:JALCOM2017}.
There are 32 crystallographic sites as shown in Table S1 of the supplementary information.
The 15 crystallographic sites for Fe atoms include seven 4{\it a} Wyckoff positions and eight 8{\it b} ones.
The Ge atoms possess 17 crystallographic sites, where two 4{\it a} sites and one 8{\it b} site are fully occupied and eleven 4{\it a} sites and three 8{\it b} sites are randomly occupied with Al atoms.
Figures\ 1(a) and 1(b) show the $a$-$c$ plane and the $a$-$b$ plane of Al$_{8.5-x}$Fe$_{23}$Ge$_{12.5+x}$, respectively.
All atoms with the 4{\it a} sites form the monolayer extending in the $b$-$c$ plane.
The atomic arrangement by the 8{\it b} sites can be regarded as a puckered two-dimensional framework also extending in the $b$-$c$ plane.
A remarkable feature of the crystal structure is the alternative stacking of the monolayer of (Al, Fe, Ge) with the 4{\it a} sites and the puckered layer of (Al, Fe, Ge) with the 8{\it b} sites along the $a$-axis.

The magnetic properties of Al$_{8.5-x}$Fe$_{23}$Ge$_{12.5+x}$ have not been investigated to our knowledge.
The Bethe-Slater curve, describing a relation between the exchange coupling and the interatomic distance, is often employed to understand the magnetism of 3$d$ transition metal elements\cite{Slater:PR1930-35,Slater:PR1930-36,Sommerfeld:book,Cardias:SR2017}.
Although Fe atom possesses a ferromagnetic (FM) ground state, a decrease of interatomic distance in a Fe-based compound can induce an antiferromagnetic (AFM) one.
In the case of a complex crystal structure, many kinds of exchange coupling between Fe atoms would invoke competition between FM and AFM states.
Al$_{8.5-x}$Fe$_{23}$Ge$_{12.5+x}$ with the lower crystal-symmetry has an opportunity of exhibiting anisotropic $x$ dependences of Fe interatomic distances, leading to a competitive relationship of two ground states.
Such an interesting phenomenon has not been well explored in a lower crystal symmetry system.
In this paper, we report the synthesis and characterization of polycrystalline samples of Al$_{8.5-x}$Fe$_{23}$Ge$_{12.5+x}$.
The magnetic and transport properties of Al$_{8.5-x}$Fe$_{23}$Ge$_{12.5+x}$ were investigated by measuring the magnetization and the electrical resistivity.

\begin{figure}
\begin{center}
\includegraphics[width=14cm]{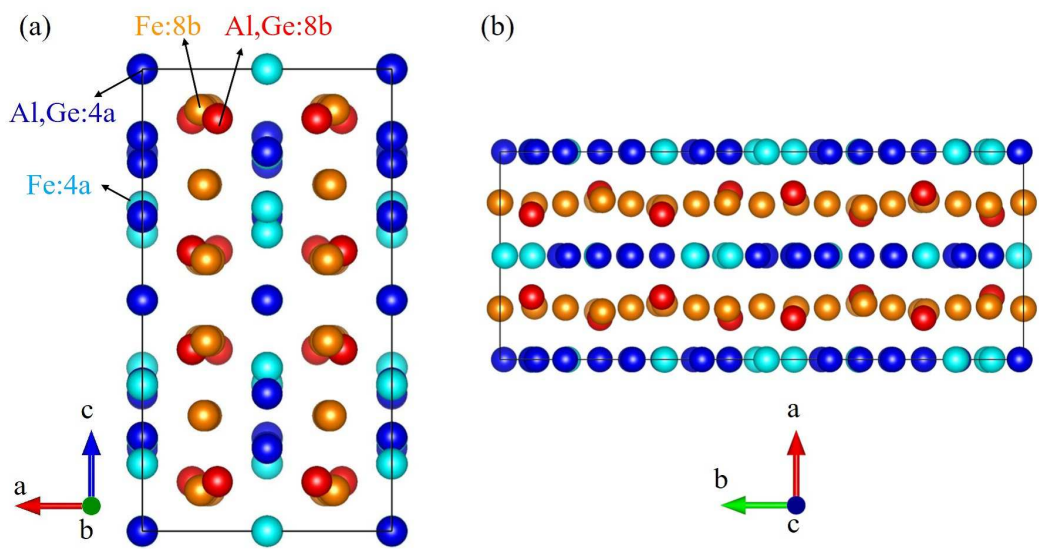}
\end{center}
\caption{Crystal structure of Al$_{8.5-x}$Fe$_{23}$Ge$_{12.5+x}$ in (a) $a$-$c$ plane and in (b) $a$-$b$ plane, respectively. The solid line represents the unit cell. The aqua, blue, orange and red balls depict Fe in the 4{\it a} sites, Al and Ge in the 4{\it a} sites, Fe in the 8{\it b} sites and Al and Ge in the 8{\it b} sites, respectively.}
\label{f1}
\end{figure}

\section{Materials and Methods}
Polycrystalline samples with different atomic compositions of Al and Ge were synthesized by an arc furnace as listed in Table 1.
The constituent elements of Al (99.99 \%), Fe (99.9 \%) and Ge (99.999 \%) were arc-melted in an Ar atmosphere on a water-cooled Cu hearth.
Each as-cast sample was placed in an evacuated quartz tube, and annealed at 800 $^{\circ}$C for 4 days and furnace cooled.

Room-temperature X-ray diffraction (XRD) patterns were collected using a Panalytical Empyrean diffractometer with Cu-K$\alpha$ radiation in Bragg-Bretano geometry.
The measured 2$\theta$ range was between 10$^{\circ}$ and 120$^{\circ}$.
The lattice parameters of each sample were obtained by the least-squares method with the help of RIETAN-FP program\cite{Izumi:SSP2007,Tsubota:SR2017}.
The metallographic examination of prepared sample was carried out by a field emission scanning electron microscope (FE-SEM; JEOL, JSM-7100F), and the atomic composition in each area ($\sim$10 $\mu$m$\times$10 $\mu$m) of the sample was determined by an energy dispersive X-ray (EDX) spectrometer coupled to the FE-SEM by averaging several data collection points.

The temperature dependence of dc magnetic susceptibility $\chi_{dc}$ ($T$) between 2 K and 300 K and the magnetization curve were measured by a Quantum Design MPMS.
The Curie temperature $T_\mathrm{C}$ and the N\'eel temperature $T_\mathrm{N}$ were determined by the minimum of temperature derivative of $\chi_{dc}$ ($T$)\cite{Oikawa:APL2001,Yu:APL2003,Kitagawa:JMMM2018,Miyahara:JSNM2018} and basically the peak position of $\chi_{dc}$ ($T$), respectively.
The temperature dependence of electrical resistivity $\rho$ ($T$) between 4 K and 300 K was measured by a dc four-probe method using a home-made system in a GM refrigerator.

\begin{table}
\caption{Starting and determined atomic compositions of main phase by EDX measurement, lattice parameters and cell volume for prepared samples.}
\label{t1}
\begin{tabular}{ccccccc}
\hline
\scriptsize{$x$} & \scriptsize{starting composition} & \scriptsize{determined composition} & \scriptsize{{\it a} (\AA)} & \scriptsize{{\it b} (\AA)} & \scriptsize{{\it c} (\AA)} & \scriptsize{{\it V} (\AA$^{3}$)}  \\
\hline
\scriptsize{0}   & \scriptsize{Al$_{19.3}$Fe$_{52.3}$Ge$_{28.4}$} & \scriptsize{Al$_{19.6(5)}$Fe$_{51.6(6)}$Ge$_{28.8(3)}$} & \scriptsize{7.9501(1)} & \scriptsize{19.7781(2)} & \scriptsize{14.6844(2)} & \scriptsize{2308.95(6)} \\
\scriptsize{0.5} & \scriptsize{Al$_{18.2}$Fe$_{52.3}$Ge$_{29.5}$} & \scriptsize{Al$_{17.9(3)}$Fe$_{51.9(5)}$Ge$_{30.2(3)}$} & \scriptsize{7.9434(1)} & \scriptsize{19.7725(2)} & \scriptsize{14.7028(2)} & \scriptsize{2309.22(5)} \\
\scriptsize{1}   & \scriptsize{Al$_{17.0}$Fe$_{52.3}$Ge$_{30.7}$} & \scriptsize{Al$_{17.3(4)}$Fe$_{51.0(3)}$Ge$_{31.7(1)}$} & \scriptsize{7.9362(1)} & \scriptsize{19.7670(2)} & \scriptsize{14.7172(2)} & \scriptsize{2308.75(5)} \\
\scriptsize{1.5} & \scriptsize{Al$_{15.9}$Fe$_{52.3}$Ge$_{31.8}$} & \scriptsize{Al$_{15.6(4)}$Fe$_{51.8(4)}$Ge$_{32.6(3)}$} & \scriptsize{7.9305(1)} & \scriptsize{19.7725(2)} & \scriptsize{14.7277(2)} & \scriptsize{2309.38(5)} \\
\scriptsize{2}   & \scriptsize{Al$_{14.8}$Fe$_{52.3}$Ge$_{32.9}$} & \scriptsize{Al$_{14.6(2)}$Fe$_{51.6(1)}$Ge$_{33.8(1)}$} & \scriptsize{7.9183(1)} & \scriptsize{19.7740(2)} & \scriptsize{14.7363(2)} & \scriptsize{2307.33(5)} \\
\scriptsize{2.5} & \scriptsize{Al$_{13.7}$Fe$_{52.3}$Ge$_{34}$}   & \scriptsize{Al$_{13.3(3)}$Fe$_{51.6(5)}$Ge$_{35.1(5)}$} & \scriptsize{7.9096(1)} & \scriptsize{19.7767(3)} & \scriptsize{14.7484(2)} & \scriptsize{2307.04(6)} \\
\scriptsize{3}   & \scriptsize{Al$_{12.5}$Fe$_{52.3}$Ge$_{35.2}$} & \scriptsize{Al$_{12.2(2)}$Fe$_{52.1(2)}$Ge$_{35.7(1)}$} & \scriptsize{7.9014(1)} & \scriptsize{19.7776(2)} & \scriptsize{14.7605(2)} & \scriptsize{2306.62(5)} \\
\hline
\end{tabular}
\end{table}

\begin{figure}
\begin{center}
\includegraphics[width=14cm]{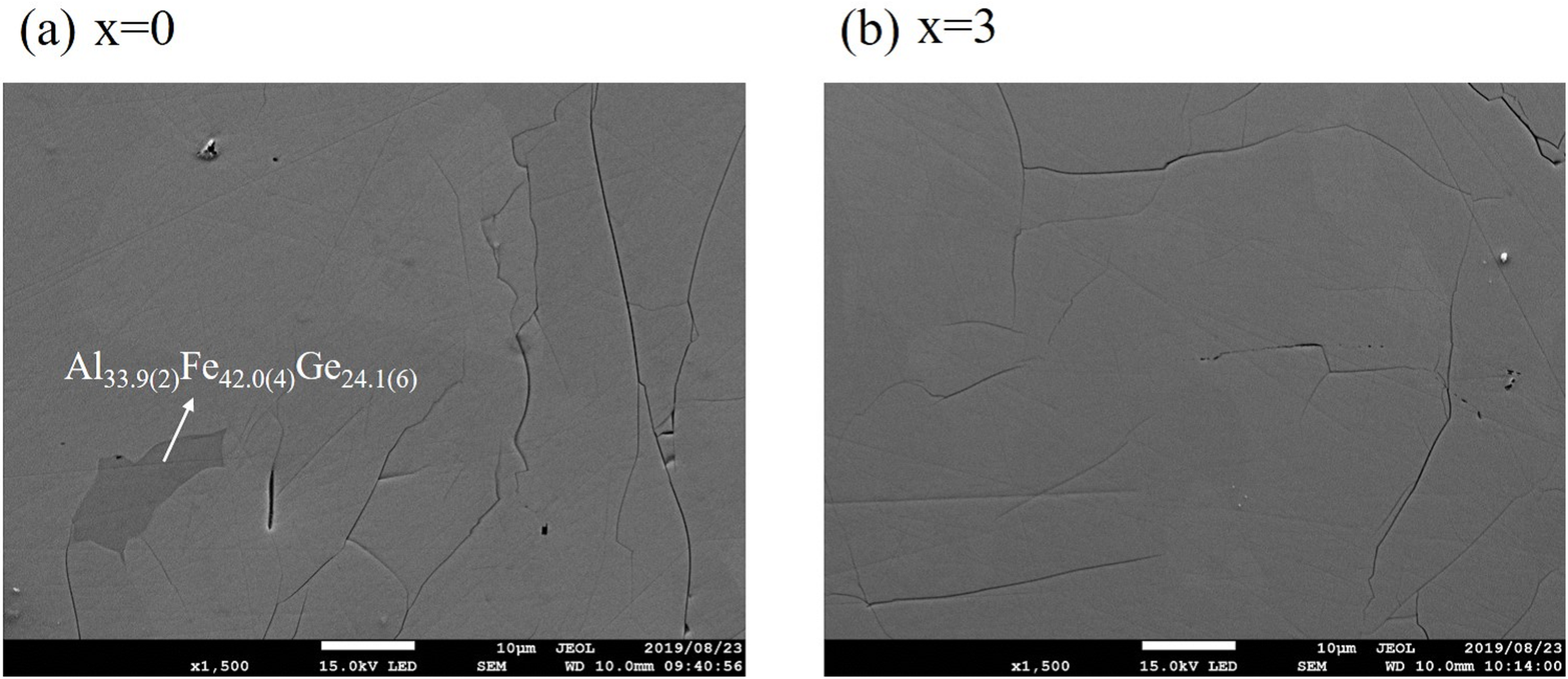}
\end{center}
\caption{Back-scattered electron images of Al$_{8.5-x}$Fe$_{23}$Ge$_{12.5+x}$ for nominal $x$ value of (a) 0 and (b) 3, respectively.}
\label{f2}
\end{figure}

\section{Results and Discussion}
\subsection{Synthesis and Crystal Structure}
The SEM images of representative samples obtained at 15 keV are shown in Fig.\ 2.
The images of the other samples are displayed in Fig.\ S1 of the supplementary information.
In each sample, no phase other than Al$_{8.5-x}$Fe$_{23}$Ge$_{12.5+x}$ could be detected, except the $x$=0 sample, where a slightly darker area appears (Fig.\ 2(a)).
The atomic composition of the secondary phase was determined to be Al$_{33.9(2)}$Fe$_{42.0(4)}$Ge$_{24.1(6)}$ by the EDX measurement.
We have confirmed that this compound is Al$_{7+x}$Fe$_{9}$Ge$_{5-x}$ reported by Reisinger et al.\cite{Reisinger:JALCOM2017} and has no magnetic phase transition down to 4 K (Fig.\ S2 in the supplementary information).
Table 1 summarizes the atomic composition of main phase determined by the EDX measurement in each sample, which agrees with the starting composition.

The example of XRD full profile is shown for $x$=1 in Fig.\ 3(a).
The diffraction peaks can be indexed with the orthorhombic {\it {\it Cmc}}2$_{1}$.
The simulated pattern calculated using the reported crystallographic parameters\cite{Reisinger:JALCOM2017} is also plotted in Fig.\ 3(a) and well matches the experimental pattern.
The full profiles of all the samples are given in Fig.\ S3 of the supplementary information.
Fig.\ 3(b) is the representative XRD patterns expanded around main peaks of the $x$=0, 1, 2 and 3 samples.
Each Miller index indicates the plane showing the strongest diffraction at the peak.
The (4 0 0) peak shifts to a higher 2$\theta$ angle with the increase of $x$, which means a contraction of the $a$-axis.
On the other hand, the (2 0 6) peak, dominated by the $l$-value connected to the $c$-axis and shifting toward a lower 2$\theta$ angle as $x$ is increased, indicates an expansion of the $c$-axis.
The weak $x$ dependence of (0 8 4) or (2 8 2) peak position with the larger $k$-value connected to the $b$-axis suggests that the $b$-axis length is insensitive to $x$.
These suggestions are confirmed in Figs.\ 4(a) to 4(c) obtained by the least-squares method.
We have also synthesized the $x$=3.5 sample, the XRD pattern of which cannot be satisfactorily explained by the orthorhombic {\it Cmc}2$_{1}$ as indicated by the arrows in Fig.\ S4 of the supplementary information.

\begin{figure}
\begin{center}
\includegraphics[width=14cm]{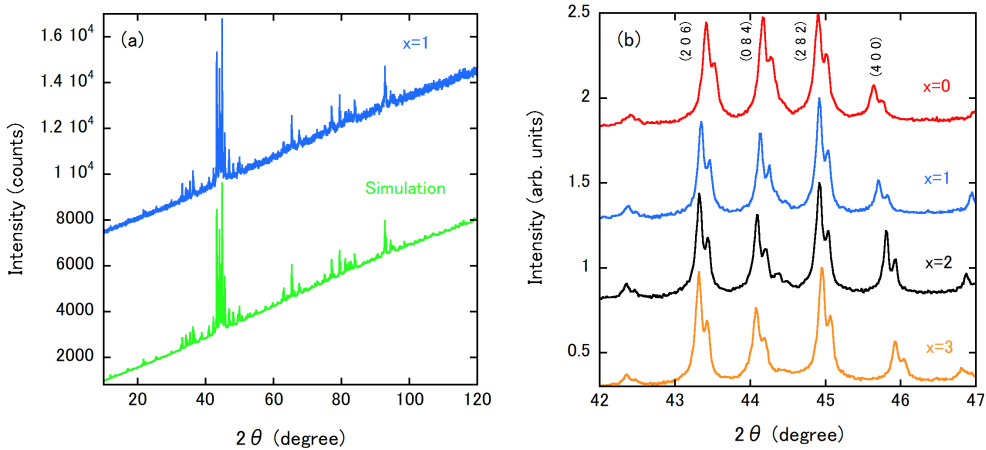}
\end{center}
\caption{(a) XRD pattern of Al$_{8.5-x}$Fe$_{23}$Ge$_{12.5+x}$ ($x$=1). The simulated pattern is also shown. (b) XRD patterns of Al$_{8.5-x}$Fe$_{23}$Ge$_{12.5+x}$ ($x$=0, 1, 2 and 3) expanded around the main peaks.}
\label{f3}
\end{figure}

\begin{figure}
\begin{center}
\includegraphics[width=11cm]{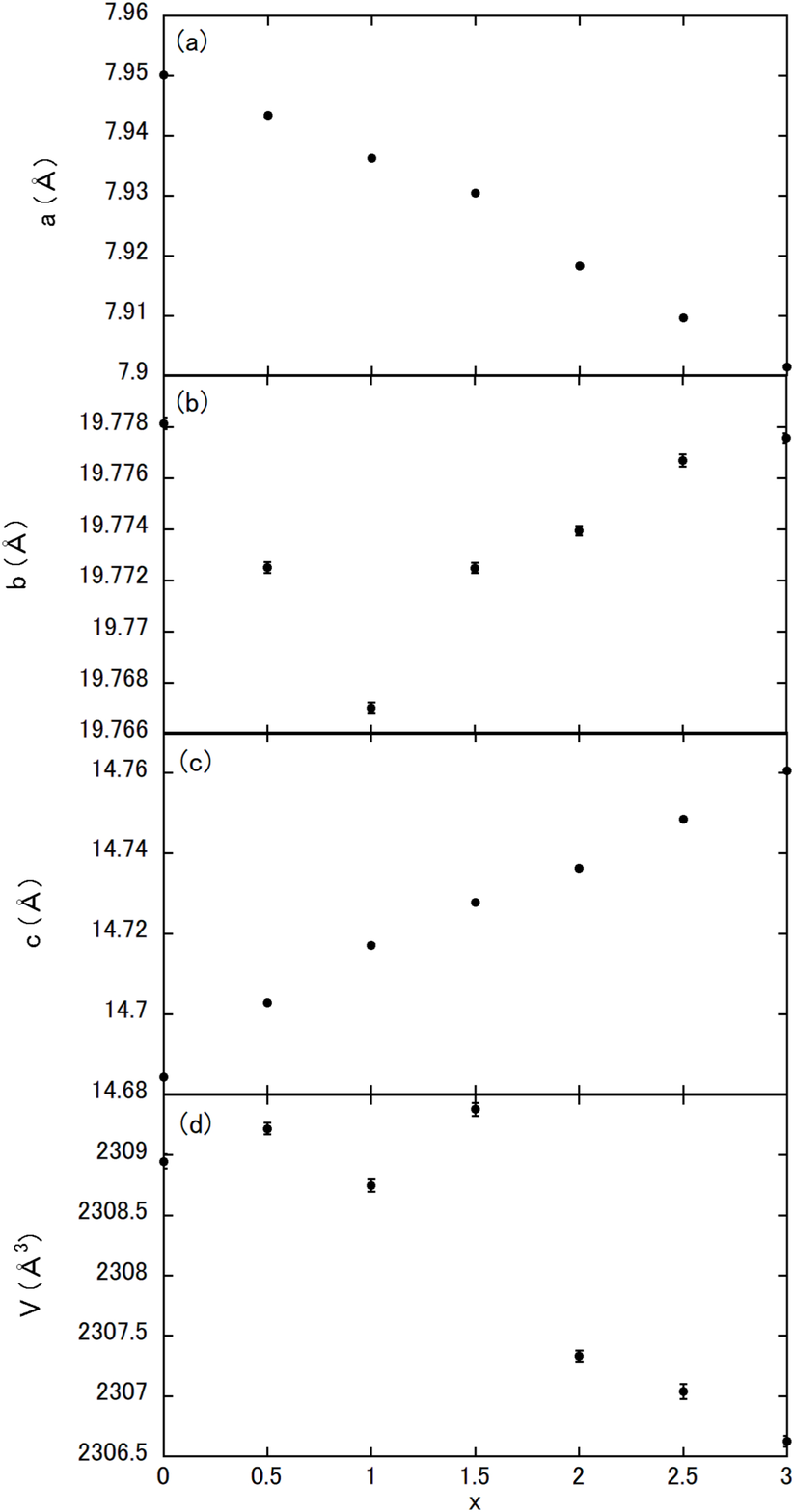}
\end{center}
\caption{$x$ dependences of (a) $a$, (b) $b$, (c) $c$ and (d) $V$, respectively, for Al$_{8.5-x}$Fe$_{23}$Ge$_{12.5+x}$ (0$\leq$$x$$\leq$3).}
\label{f4}
\end{figure}

As explained in the Introduction, the Fe-Fe bond length would play an important role in determining the magnetic ground state of Al$_{8.5-x}$Fe$_{23}$Ge$_{12.5+x}$.
Based on the atomic positions reported by Reisinger et al.\cite{Reisinger:JALCOM2017}, the schematic view of Fe-Fe bonds are exhibited in Figs.\ 5 and 6 for $x$$\sim$3 (see also Table S2 of the supplementary information).
Figures 5(a) and 5(b) demonstrate the Fe-Fe bonds with lengths less than 2.56 \AA.
The numbers of Fe atoms correspond to those listed in Table S1, and the aqua and orange balls depict the Fe atoms with the 4{\it a} sites (Fe:4{\it a}, Fe1$\sim$Fe7) and the 8{\it b} sites (Fe:8{\it b}, Fe8$\sim$Fe15), respectively.
As can be seen from Fig.\ 5(a), Fe8, Fe9, Fe10 and Fe13 atoms form a linear chain along the $b$-axis with rather short interatomic distances (see also Fig.\ 7(a)).
Another major bonds described in Fig.\ 5(b) can be regarded that Fe:4{\it a} atoms bridge the two neighboring puckered two-dimensional layers by Fe:8{\it b} atoms.
Figures 6(a) and 6(b) draw the Fe-Fe bonds with interatomic distances between 2.56 \AA \hspace{1.5mm} and 3 \AA.
The notable bonds in Fig.\ 6(a) are orange ones extending along the $c$-axis (see also Fig.\ 7(c)).
The view in Fig.\ 6(b) (see also Fig.\ 7(d)) suggests that Fe2-Fe8 (Fe10), Fe3-Fe14 (F15), Fe4-Fe10 (Fe13), Fe5-Fe9 (Fe13) and Fe7-Fe14 (Fe15) bonds play the same role as do the bonds in Fig.\ 5(b).

The impacts of replacement of Al atom with Ge atom on Fe-Fe bond lengths selected in Figs.\ 5 and 6 are investigated as the bond length vs. $x$ plots in Figs.\ 7(a) to 7(d).
The bond lengths are calculated employing the atomic positions reported by Reisinger et al.\cite{Reisinger:JALCOM2017}
The direction of Fe-Fe bonds in Fig.\ 7(a) is parallel to the $b$-axis (see also Fig.\ 5(a)), which is rather insensitive to the change of $x$.
All Fe:4{\it a}-Fe:8{\it b} bonds steadily shrink with increasing $x$, mainly due to the decreasing $a$-axis (see Figs.\ 7(b) and 7(d)).
Contrasted to the result, the Fe:8{\it b}-Fe:8{\it b} bonds running along the $c$-axis systematically expand with the increase of $x$.

\begin{figure}
\begin{center}
\includegraphics[width=14cm]{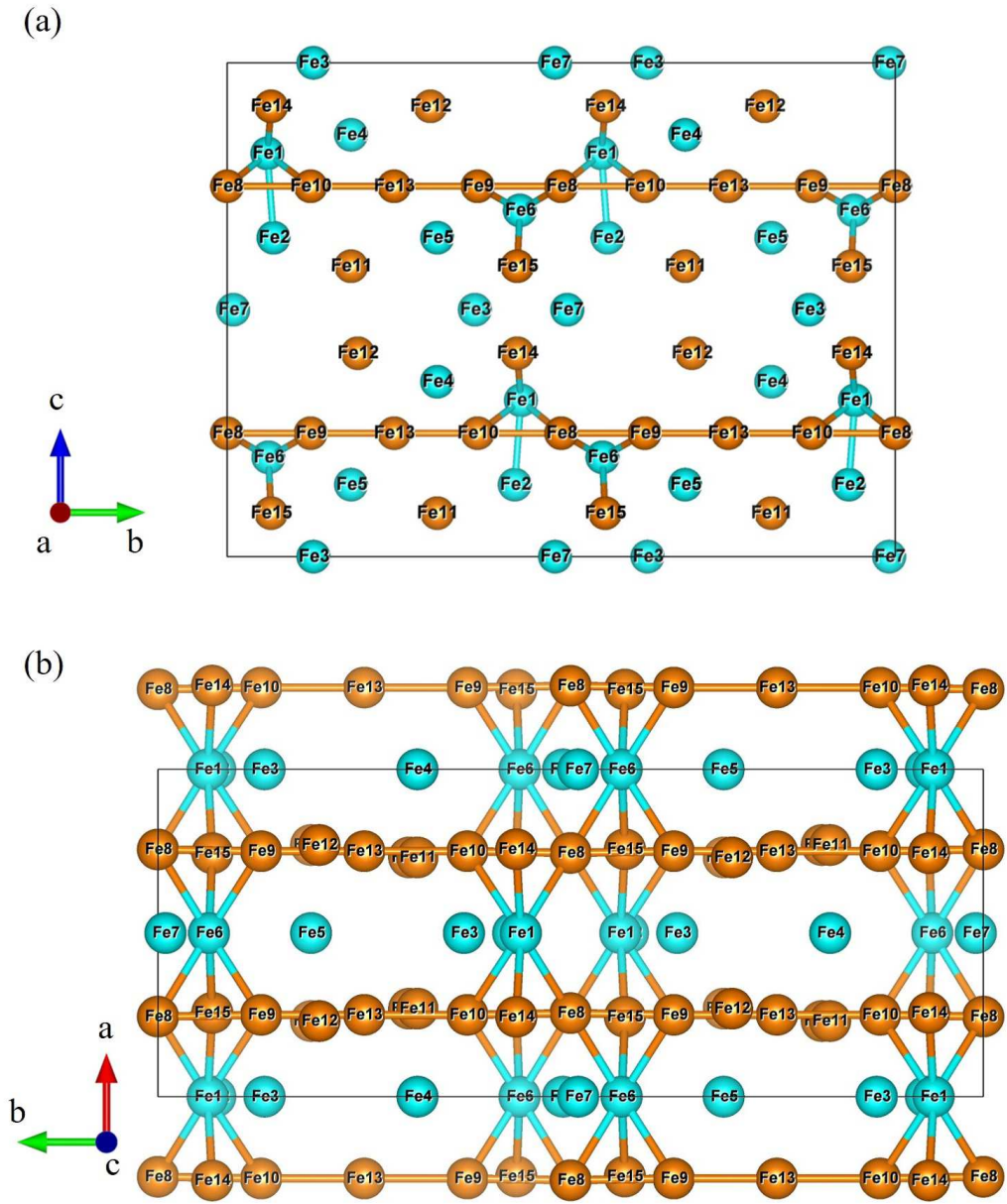}
\end{center}
\caption{Schematic view of Fe-Fe bonds with lengths less than 2.56 \AA \hspace{1.5mm} in (a) $b$-$c$ plane and in (b) $a$-$b$ plane, respectively, for $x$=3 sample. The numbers of Fe atoms correspond to those listed in Table S1 of the supplementary information. The solid line represents the unit cell.}
\label{f5}
\end{figure}

\begin{figure}
\begin{center}
\includegraphics[width=14cm]{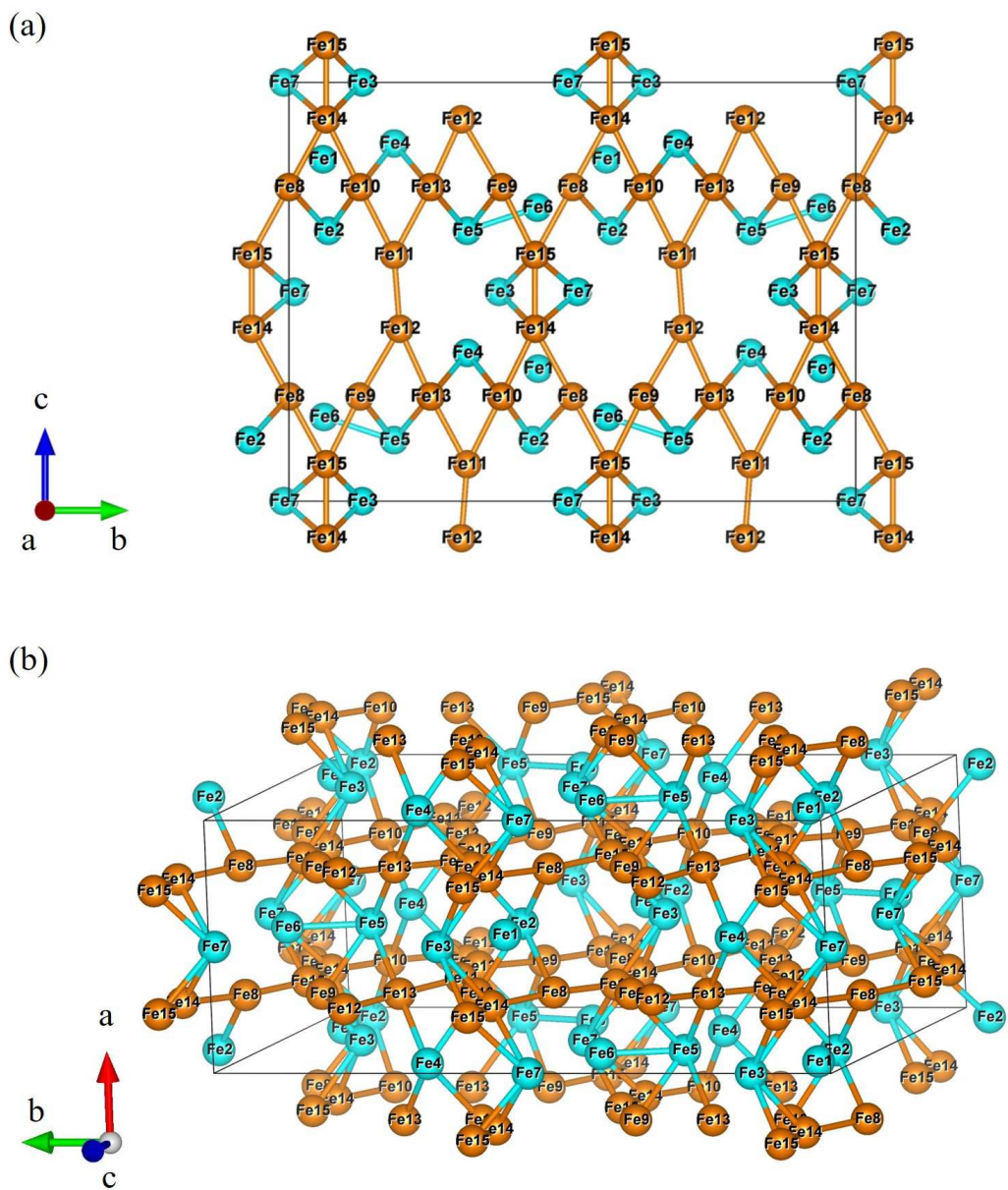}
\end{center}
\caption{Schematic view of Fe-Fe bonds with lengths between 2.56 \AA \hspace{1.5mm} and 3 \AA \hspace{1.5mm} in (a) $b$-$c$ plane and in (b) crystal structure tilted from $a$-$b$ plane, respectively, for $x$=3 sample. The numbers of Fe atoms correspond to those listed in Table S1 of the supplementary information. The solid line represents the unit cell.}
\label{f6}
\end{figure}

\begin{figure}
\begin{center}
\includegraphics[width=15cm]{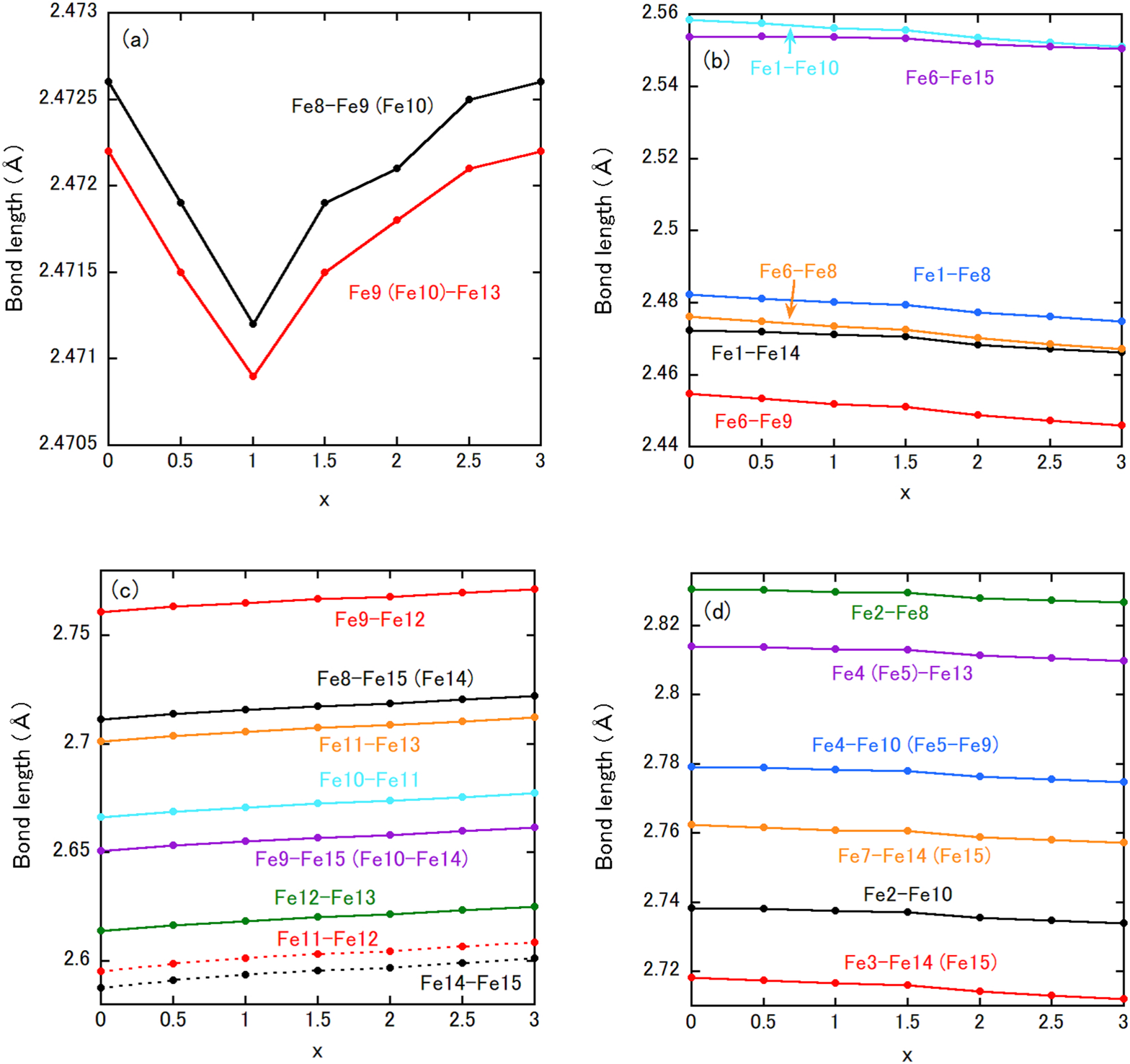}
\end{center}
\caption{Fe-Fe bond length vs $x$ plots in Al$_{8.5-x}$Fe$_{23}$Ge$_{12.5+x}$ (0$\leq$$x$$\leq$3) for (a) Fe:8{\it b}-Fe:8{\it b} with length less than 2.56 \AA, (b) Fe:4{\it a}-Fe:8{\it b} less than 2.56 \AA, (c) Fe:8{\it b}-Fe:8{\it b} between 2.56 and 3 \AA \hspace{1.5mm} and (d) Fe:4{\it a}-Fe:8{\it b} between 2.56 and 3 \AA, respectively.}
\label{f7}
\end{figure}

\subsection{Magnetic Susceptibility}
Figures 8(a) and 8(b) show $\chi_{dc}$ ($T$) under the external field $H$ of 10 Oe for all the samples and those for the $x$=2.5 and 3 samples with the expanded view, respectively, and $T_\mathrm{C}$ and $T_\mathrm{N}$ are collected in Table 2.
In the inset of Fig.\ 8(b), the low-temperature part of $\chi_{dc}$ ($T$) of the $x$=1 sample is given.
Each sample in $x$=0 to 1 exhibits the typical $\chi_{dc}$ ($T$) behavior for a ferromagnet, characterized by the steep increase of $\chi_{dc}$ below $T_\mathrm{C}$.
The obvious ferromagnetic behavior is not observed in the $x$=2.5 or 3 sample (Fig.\ 8(b)).
For each sample, the sharp peak indicates a signature of AFM ordering, which is consistent with the results of magnetization curves as mentioned below.
In the $x$=1.5 ($x$=2) sample, an abrupt drop of $\chi_{dc}$ is observed below approximately 85 K (110 K), that is reminiscent of the FM-AFM transitions as reported in FeRh$_{1-x}$Pt$_{x}$, Ce(Fe$_{1-x}$Co$_{x}$)$_{2}$, Mn$_{3}$GaC, Mn$_{1-x}$Cr$_{x}$CoGe and so on\cite{Yuasa:JPSJ1994,Rastogi:book,Bouchaud:JAP1966,Trung:APL2010}.
The paramagnetic-FM and FM-AFM transitions are more discernible in the temperature dependences of ac magnetization (Fig.\ S2 in the supplementary information).
The entering into AFM ordering on cooling in the $x$=1.5 or 2 sample is also supported by $\rho$ ($T$) (Fig.\ 11).
The AFM ordering probably occurs also in the $x$=1 sample, showing a small hump at approximately 35 K (see the inset of Fig.\ 8(b)).

The temperature dependences of inverse $\chi_{dc}$ for all the samples would obey the Curie-Weiss law above $T_\mathrm{C}$ or $T_\mathrm{N}$ as indicated by the solid lines (Fig.\ 9).
The effective magnetic moment $\mu_\mathrm{eff}$ and the Weiss temperature are summarized in Table 2.
$\mu_\mathrm{eff}$ is near to theoretical one associated with Fe$^{2+}$ (4.90 $\mu_\mathrm{B}$/Fe) or Fe$^{3+}$ (5.92 $\mu_\mathrm{B}$/Fe) ion, which means the well localized nature of Fe moment in the whole composition range.
Even in the $x$=2.5 and 3 samples with the AFM ground states, the Weiss temperatures are positive, that implies a dominance of FM interaction between Fe magnetic moments.
We note that FeRh$_{1-x}$Pt$_{x}$ with the composition showing only the AFM ordering also possesses a positive Weiss temperature\cite{Yuasa:JPSJ1994}.
It should be additionally commented that a helical antiferromagnet ZnCr$_{2}$Se$_{4}$ shows a positive Weiss temperature\cite{Hemberger:PRL2007}.
The reason is that the complex AFM magnetic structure is established through competing FM and AFM interactions between Cr atoms and the Weiss temperature is obtained by averaging these interactions with dominating FM one.
As mentioned below, the present compound would possess competing FM and AFM interactions, resulting in a complex magnetic structure.
Therefore we infer that the origin of positive Weiss temperature even in the AFM state is akin to that in ZnCr$_{2}$Se$_{4}$.

\begin{figure}
\begin{center}
\includegraphics[width=14cm]{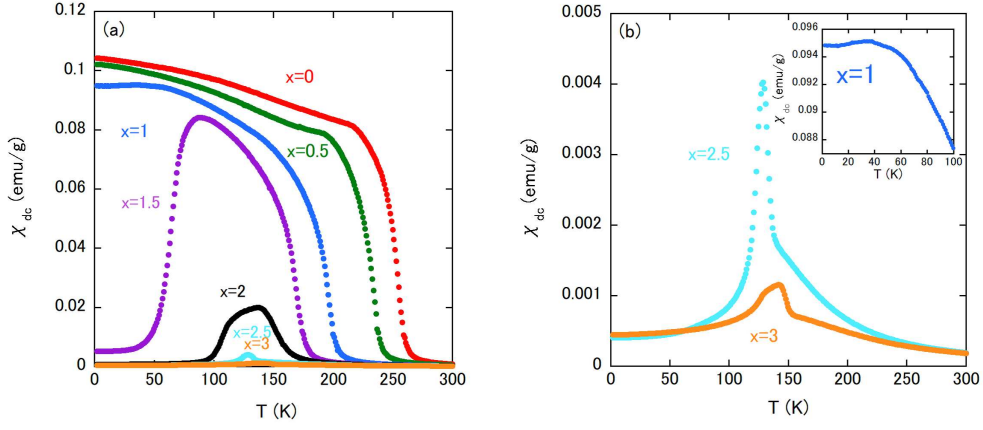}
\end{center}
\caption{(a) Temperature dependences of $\chi_{dc}$ of Al$_{8.5-x}$Fe$_{23}$Ge$_{12.5+x}$ (0$\leq$$x$$\leq$3) under external field of 10 Oe. (b) Expanded view of $\chi_{dc}$ ($T$) for $x$=2.5 and 3 samples. The inset is low temperature $\chi_{dc}$ ($T$) of the $x$=1 sample.}
\label{f8}
\end{figure}

\begin{figure}
\begin{center}
\includegraphics[width=8cm]{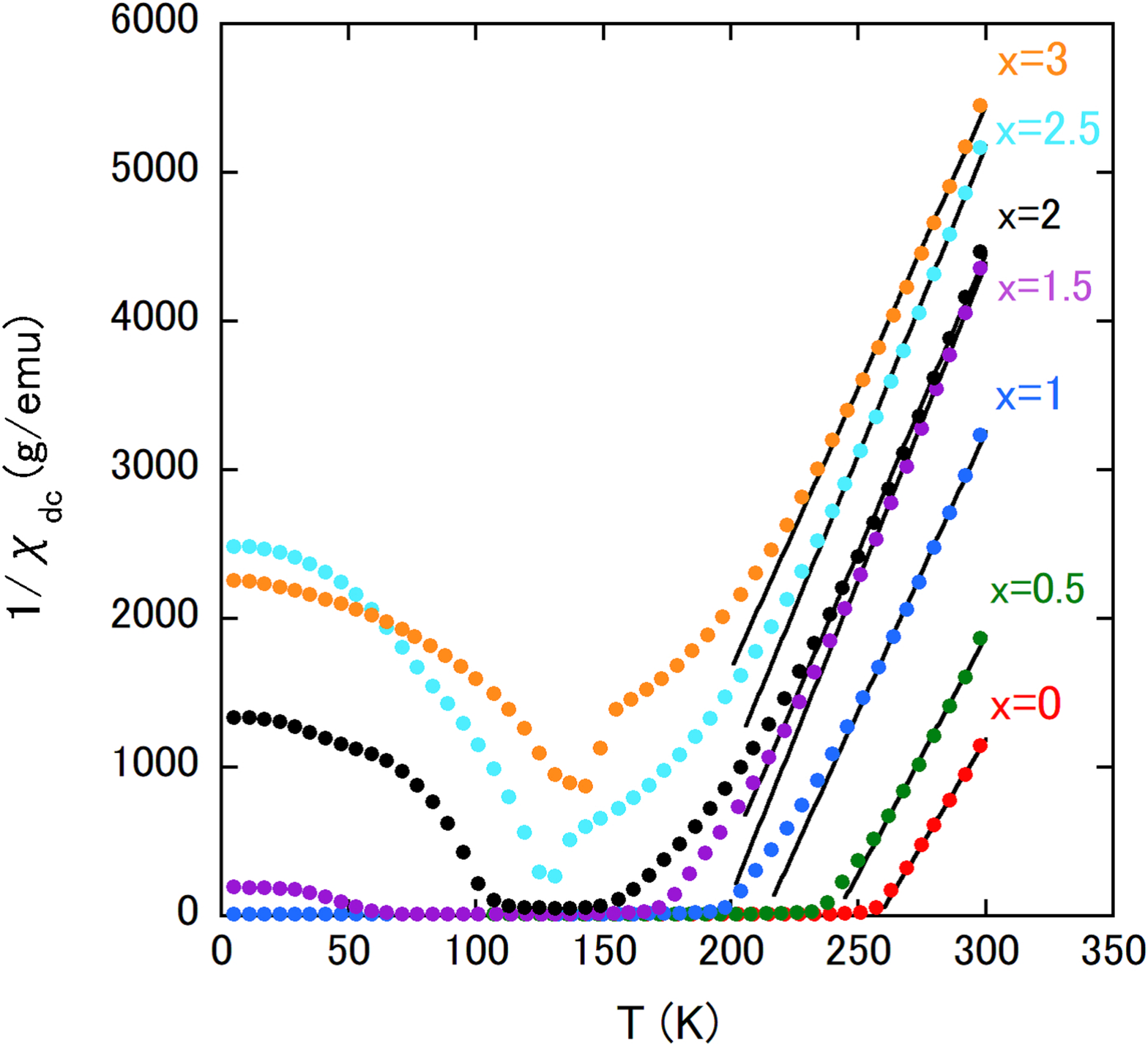}
\end{center}
\caption{The inverse  $\chi_{dc}$ ($T$) of Al$_{8.5-x}$Fe$_{23}$Ge$_{12.5+x}$ (0$\leq$$x$$\leq$3).}
\label{f9}
\end{figure}

\begin{table}
\caption{$T_\mathrm{C}$, $T_\mathrm{N}$, $\mu_\mathrm{eff}$, Weiss temperature, $M$ at 2 K and 50 kOe, $H_{m}$ showing metamagnetic-like transition at 2 K and $\rho$ at room temperature of Al$_{8.5-x}$Fe$_{23}$Ge$_{12.5+x}$ (0$\leq$$x$$\leq$3).}
\label{t2}
\begin{tabular}{cccccccc}
\hline
\footnotesize{$x$} & \footnotesize{$T_\mathrm{C}$ (K)} & \footnotesize{$T_\mathrm{N}$ (K)} & \footnotesize{$\mu_\mathrm{eff}$ ($\mu_\mathrm{B}$/Fe)} & \footnotesize{Weiss} & \footnotesize{$M$ ($\mu_\mathrm{B}$/Fe)} & \footnotesize{$H_{m}$ (kOe)} & \footnotesize{$\rho$ (RT)}  \\
 & & & & \footnotesize{temperature (K)} & & & \footnotesize{($\mu\Omega$cm)} \\
\hline
\footnotesize{0}   & \footnotesize{255} & \footnotesize{-}  & \footnotesize{5.47} & \footnotesize{258} & \footnotesize{0.96} & \footnotesize{-} & \footnotesize{236} \\
\footnotesize{0.5} & \footnotesize{236} & \footnotesize{-}  & \footnotesize{5.16} & \footnotesize{241} & \footnotesize{1.00} & \footnotesize{-} & \footnotesize{433} \\
\footnotesize{1}   & \footnotesize{196} &  \footnotesize{36} & \footnotesize{4.79} & \footnotesize{212} & \footnotesize{0.93} & \footnotesize{1} & \footnotesize{480} \\
\footnotesize{1.5} & \footnotesize{169} &  \footnotesize{85} & \footnotesize{4.52} & \footnotesize{197} & \footnotesize{0.60} & \footnotesize{14} & \footnotesize{726} \\
\footnotesize{2}   & \footnotesize{152} & \footnotesize{107} & \footnotesize{4.67} & \footnotesize{189} & \footnotesize{0.52} & \footnotesize{18} & \footnotesize{820} \\
\footnotesize{2.5} & \footnotesize{-}  & \footnotesize{129} & \footnotesize{4.62} & \footnotesize{175} & \footnotesize{0.30} & \footnotesize{20} & \footnotesize{613} \\
\footnotesize{3}   &  \footnotesize{-}  & \footnotesize{143} & \footnotesize{4.85} & \footnotesize{156} & \footnotesize{0.18} & \footnotesize{26} & \footnotesize{505} \\
\hline
\end{tabular}
\end{table}

\subsection{Magnetization curve}
Figures 10(a) to 10(d) show the isothermal $M$-$H$ ($M$: magnetization) curves measured at several temperatures denoted in the figures for (a) $x$=0, (b)(c) $x$=1.5 and (d) $x$=3, respectively.
The data of the other samples are summarized in the supplementary information (Figs.\ S5 to S8).
A soft ferromagnetic behavior is confirmed in the  $x=$0 sample.
The noticeable feature is, for example at 2 K, the rather low value of $M$ at 50 kOe compared to $\mu_\mathrm{eff}$, which means a possible complex magnetic structure.
In the $x=$1.5 sample, below approximately $T_\mathrm{C}$ (=169 K), $M$ steeply increases at the initial magnetization process, which is consistent with the FM state (Fig.\ 10(b)).
On the contrary, as shown in Fig.\ 10(c), such a steep increase of $M$ is not observed below $T_\mathrm{N}$ (=85 K).
The initial magnetization process at 2 K is particularly similar to that of an AFM state.
It is interesting that even in the AFM state, the irreversibility of magnetization curve becomes more obvious with decreasing temperature.
This might suggest a possible coexistence of AFM and FM states and further study is needed to clarify the origin.
For the $x=$3 sample (Fig.\ 10(d)), $M$-$H$ curves at 100 K or 2 K below $T_\mathrm{N}$ (=143 K) shows a metamagnetic-like behavior at $H$=20$\sim$30 kOe, which is consistent with the AFM ground state.
Contrasted with $M$-$H$ curves of the $x$=1.5 sample below $T_\mathrm{N}$, no irreversibility is detected.

\begin{figure}
\begin{center}
\includegraphics[width=14cm]{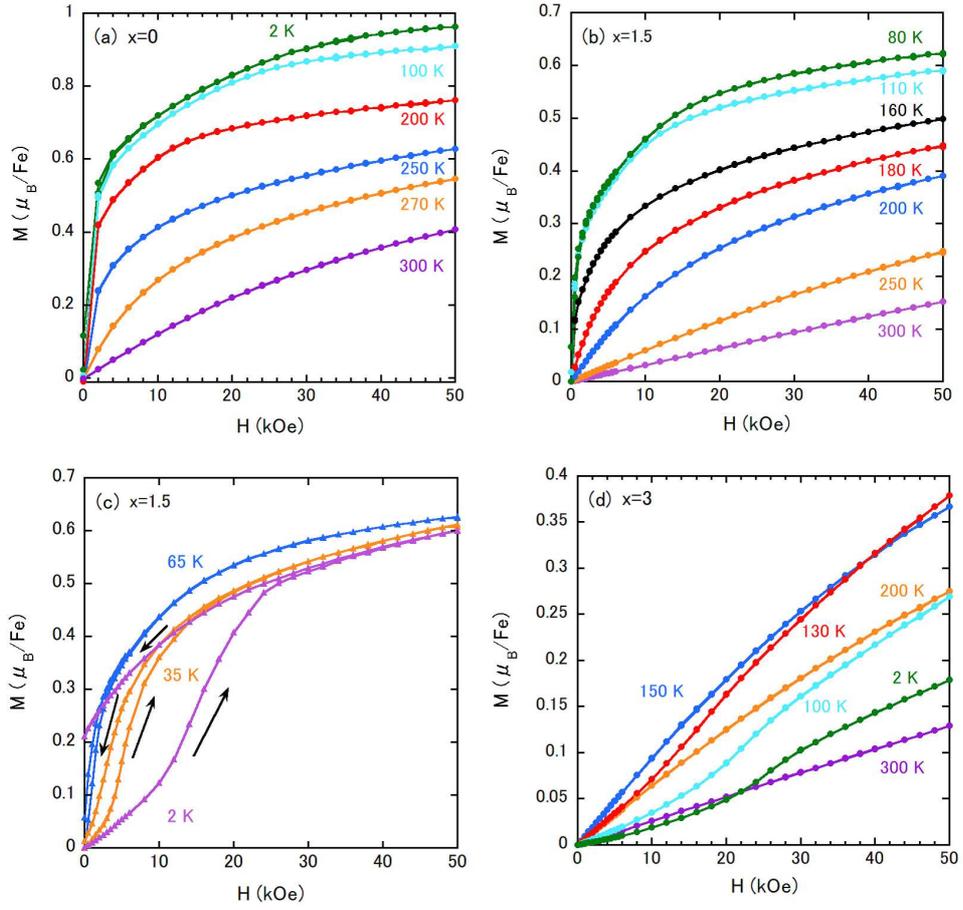}
\end{center}
\caption{$M$-$H$ curves measured at several temperatures denoted in figures for Al$_{8.5-x}$Fe$_{23}$Ge$_{12.5+x}$ with (a) $x$=0, (b)(c) $x$=1.5 and (d) $x$=3, respectively.}
\label{f10}
\end{figure}

\begin{figure}
\begin{center}
\includegraphics[width=8cm]{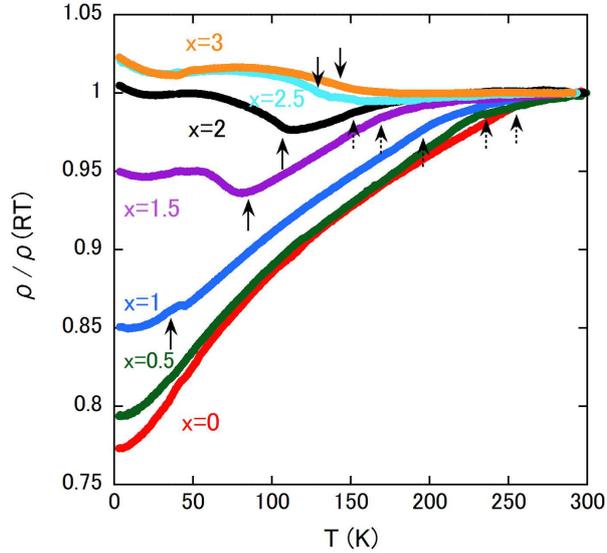}
\end{center}
\caption{Temperature dependences of $\rho$ for Al$_{8.5-x}$Fe$_{23}$Ge$_{12.5+x}$ (0$\leq$$x$$\leq$3). For each sample, $\rho$ ($T$) is normalized by the room temperature value listed in Table 2.}
\label{f11}
\end{figure}

\subsection{Electrical Resistivity}
$\rho$ ($T$) of all the samples are exhibited in Fig.\ 11, where each $\rho$ ($T$) is normalized by the room-temperature value listed in Table 2.
The room-temperature values indicate the metallic state in all samples.
The temperature variation is basically small due to the atomic disorder between Al and Ge atoms.
$T_\mathrm{C}$ and $T_\mathrm{N}$ determined by $\chi_{dc}$ ($T$) measurements are indicated by the dotted and solid arrows, respectively.
In $x$=1.5 to 3 samples showing obvious AFM orderings, each $\rho$ exhibits an upturn below approximately $T_\mathrm{N}$, which suggests a partial gap-opening at the Fermi level.
We note here that, in the inherent slab-like structure of the present system, the decrease of $a$-axis and increase of $c$-axis with increasing $x$ lead to an enhanced two-dimensional nature of crystal, which might be related with the instability of Fermi surface below $T_\mathrm{N}$.

\subsection{Magnetic Phase Diagram}
Shown in Fig.\ 12 is the magnetic phase diagram of Al$_{8.5-x}$Fe$_{23}$Ge$_{12.5+x}$ (0$\leq$$x$$\leq$3), which clearly demonstrates the competition between FM and AFM states.
Following the fact that, in the Th$_{2}$Ni$_{17}$-type RE$_{2}$Fe$_{17}$ (RE:rare earth) with four different Wyckoff positions of Fe atoms, the Fe-Fe bonds with lengths below and above 2.5 \AA \hspace{1.5mm} possess the AFM and the FM exchange interactions, respectively\cite{Givord:IEEE1974}, those shown in Figs.\ 7(a) and 7(b) (Figs.\ 7(c) and 7(d)) for Al$_{8.5-x}$Fe$_{23}$Ge$_{12.5+x}$ would be in the range of AFM (FM) coupling.
Thus in the present compound, the FM and AFM exchange interactions would compete with each other.
As considered in the subsection of Synthesis and Crystal Structure, the bond lengths of Fe:4{\it a}-Fe:8{\it b} and those of Fe:8{\it b}-Fe:8{\it b} along the $c$-axis decrease and increase with increasing $x$, respectively.
The Fe:8{\it b}-Fe:8{\it b} bonds along the $b$-axis, which would have the strongest AFM coupling, are rather insensitive to $x$ value.
So we speculate that, as $x$ is increased, the FM interaction due to Fe:8{\it b}-Fe:8{\it b} bonds shown in Fig.\ 7(c) may be weakened by the expansion of length, and on the contrary the AFM one due to Fe:4{\it a}-Fe:8{\it b} in Fig.\ 7(b) is enhanced by the shrinkage of length.
This would cause the dominance of AFM ground state with the increment of $x$, which is also supported by the cell volume being smaller especially at $x$$\geq$1.5 as shown in Fig.\ 4(d).

In the intermediate range of $x$ (1$\leq$$x$$\leq$2), the FM-AFM phase transition is observed on cooling.
The temperature dependence of $\chi_{dc}$ and that of $\rho$ characterized by the Fermi surface instability observed as the upturn of $\rho$ below $T_\mathrm{N}$ are similar to those of several compounds such as FeRh$_{1-x}$Pt$_{x}$ and Ce(Fe$_{1-x}$Co$_{x}$)$_{2}$, which also show the FM-AFM transitions\cite{Yuasa:JPSJ1994,Fukuda:PRB2001}.
It is peculiar that Al$_{8.5-x}$Fe$_{23}$Ge$_{12.5+x}$ possesses the complex crystal structure and the complexity may be a new route for observing the FM-AFM transition.

\begin{figure}
\begin{center}
\includegraphics[width=8cm]{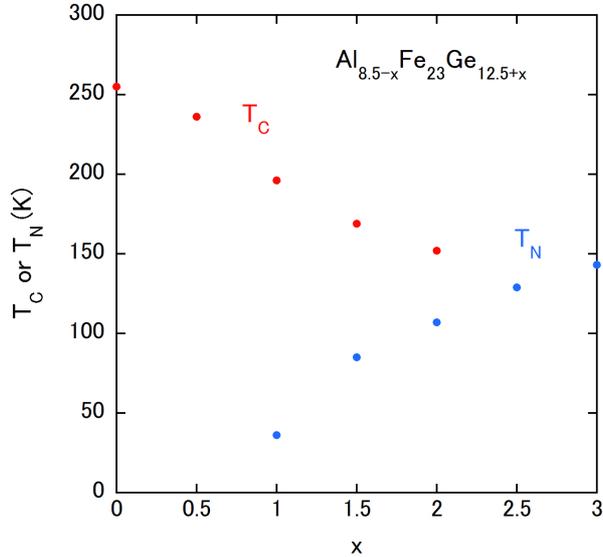}
\end{center}
\caption{Magnetic phase diagram of Al$_{8.5-x}$Fe$_{23}$Ge$_{12.5+x}$ (0$\leq$$x$$\leq$3).}
\label{f12}
\end{figure}

\section{Summary}
Solid solutions of polycrystalline Al$_{8.5-x}$Fe$_{23}$Ge$_{12.5+x}$ with the orthorhombic structure were prepared in the range of 0$\leq$$x$$\leq$3.
The compound shows the anisotropic change of lattice parameters; decreasing (increasing) $a$ ($c$) with increasing $x$ and $b$ with almost insensitive to $x$.
The magnetic properties have revealed the competition between the FM and AFM states, depending on the $x$ value.
Starting from $x$=0, $T_\mathrm{C}$ is suppressed and disappears at the $x$=2.5 sample, which instead shows the AFM transition.
The $T_\mathrm{N}$ seems to appear at $x$=1 and grows to 143 K at $x$=3.
In the intermediate range of $x$ (1$\leq$$x$$\leq$2), the FM-AFM phase transition occurs on cooling.
The AFM phase can be characterized by $\rho$ ($T$) showing the upturn below $T_\mathrm{N}$ accompanying a Fermi surface instability.
The instability would be ascribed to the enhancement of two-dimensionality of crystal structure.
The anisotropic $x$ dependences of Fe-Fe lengths, originating from the complex low-symmetry crystal structure, would lead to the competition between the  FM and AFM states.

\section*{Acknowledgments}
J.K. is grateful for the support provided by the Comprehensive Research Organization of Fukuoka Institute of Technology.


\begin{thebibliography}{99}
\bibitem{Takeda:JPSP1995}
N. Takeda, J. Kitagawa, M. Ishikawa, J. Phys. Soc. Jpn. {\bf 64} (1995) 387-390.
\bibitem{Kitagawa:PRB1996}
J. Kitagawa, N. Takeda, M. Ishikawa, Phys. Rev. B {\bf 53} (1996) 5101-5103.

\bibitem{Kitagawa:JAC1997}
J. Kitagawa, N. Takeda, M. Ishikawa, J. Alloys Compd. {\bf 256} (1997) 48-56.

\bibitem{Kitagawa:PRB1998}
J. Kitagawa, N. Takeda, M. Ishikawa, T. Yoshida, A. Ishiguro, N. Kimura, T. Komatsubara, Phys. Rev. B {\bf 57} (1998) 7450-7453.

\bibitem{Kitagawa:JPSJ2000}
J. Kitagawa, N. Takeda, M. Ishikawa, M. Nakayama, N. Kimura, T. Komatsubara, J. Phys. Soc. Jpn. {\bf 69} (2000) 883-887.

\bibitem{Custers:NM2012}
J. Custers, K.-A. Lorenzer, M. M\"uller, A. Prokofiev, A. Sidorenko, H. Winkler, A. M. Strydom, Y. Shimura, T. Sakakibara, R. Yu, Q. Si, S. Paschen, Nat. Mater. {\bf 11} (2012) 189-194.

\bibitem{Fujita:PRB2003}
A. Fujita, S. Fujieda, Y. Hasegawa, K. Fukamichi, Phys. Rev. B {\bf 67} (2003) 104416.

\bibitem{Tawara:JJAP1973}
Y. Tawara, H. Senno, Jpn. J. Appl. Phys. {\bf 12} (1973) 761-762.

\bibitem{Sagawa:JAP1984}
M. Sagawa, S. Fujimura, N. Togawa, H. Yamamoto, Y. Matsuura, J. Appl. Phys. {\bf 55} (1984) 2083-2087.

\bibitem{Croaf:JAP1984}
J. J. Croat, J. F. Herbst, R. W. Lee, F. E. Pinkerton, J. Appl. Phys. {\bf 55} (1984) 2078-2082..

\bibitem{Coey:JMMM1990}
J. M. D. Coey, H. Sun, J. Magn. Magn. Mater. {\bf 87} (1990) L251-L254..

\bibitem{Brown:ChemMater2006}
S. R. Brown, S. M. Kauzlarich, F. Gascoin, G. J. Snyder, Chem. Mater. {\bf 18} (2006) 1873-1877.

\bibitem{Reisinger:JALCOM2017}
 G. R. Reisinger, H. S. Effenberger, K. W. Richter, J. Alloys Compd. 693 (2017) 692-699.

\bibitem{Slater:PR1930-35}
J. C. Slater, Phys. Rev. {\bf 35} (1930) 509-529.

\bibitem{Slater:PR1930-36}
J. C. Slater, Phys. Rev. {\bf 36} (1930) 57-64.

\bibitem{Sommerfeld:book}
A. Sommerfeld and H. Beter, Handbuch der physik, Springer, Berlin, {\bf 24} (1933).

\bibitem{Cardias:SR2017}
R. Cardias, A. Szilva, A. Bergman, I. Di Marco, M. I. Katsnelson, A. I. Lichtenstein, L. Nordstr\"{o}m, A. B. Klautau, O. Eriksson, Y. O. Kvashnin, Sci. Rep. {\bf 7} (2017) 4058.

\bibitem{Izumi:SSP2007}
F. Izumi, K. Momma, Solid State Phenom. {\bf 130} (2007) 15-20.

\bibitem{Tsubota:SR2017}
M. Tsubota, J. Kitagawa, Sci. Rep. {\bf 7} (2017) 15381.

\bibitem{Oikawa:APL2001}
K. Oikawa, L. Wulff, T. Iijima, F. Gejima, T. Ohmori, A. Fujita, K. Fukamichi, R. Kainuma, K. Ishida, Appl. Phys. Lett. {\bf 79} (2001) 3290-3292.

\bibitem{Yu:APL2003}
M. -H. Yu, L. H. Lewis, A. R. Moodenbaugh, J. Appl. Phys. {\bf 93} (2003) 10128-10130..

\bibitem{Kitagawa:JMMM2018}
J. Kitagawa, K. Sakaguchi, J. Magn. Magn. Mater. {\bf 468} (2018) 115-122.

\bibitem{Miyahara:JSNM2018}
J. Miyahara, N, Shirakawa, Y. Setoguchi, M. Tsubota, K. Kuroiwa, J. Kitagawa, J. Supercond. Nov. Magn. {\bf 31} (2018) 3559-3564.

\bibitem{Yuasa:JPSJ1994}
S. Yuasa, H. Miyajima, Y. Otani, J. Phys. Soc. Jpn. {\bf 63} (1994) 3129-3144.

\bibitem{Rastogi:book}
A. K. Rastogi, A. P. Murani,  Theoretical and Experimental Aspects of Valence Fluctuations and Heauy Fermions, Plenum, New York, (1987).

\bibitem{Bouchaud:JAP1966}
J. -P. Bouchaud, R. Fruchart, R. Pauthenet, M. Guillot, H. Bartholin, F. Chaiss\'{e}, J. Appl. Phys. {\bf 37} (1966) 971-972.

\bibitem{Trung:APL2010}
N. T. Trung, V. Biharie, L. Zhang, L. Caron, K. H. J. Buschow, E. Br\"{u}ck, Appl. Phys. Lett. {\bf 96} (2010) 162507.

\bibitem{Hemberger:PRL2007}
J. Hemberger, H. -A. Krug von Nidda, V. Tsurkan, A. Loidl, Phys. Rev. Lett. {\bf 98} (2007) 147203.

\bibitem{Givord:IEEE1974}
D. Givord, R. Lemaire, IEEE Trans. Mag. {\bf MAG-10} (1974) 109-113.

\bibitem{Fukuda:PRB2001}
H. Fukuda, H. Fujii, H. Kamura, Y. Hasegawa, T. Ekino, N. Kikugawa, T. Suzuki, T. Fujita, Phys. Rev. B {\bf 63} (2001) 054405.

\end{thebibliography}
\end{document}